 \documentclass[authoryear,preprint,review,12pt]{elsarticle}

\usepackage{amssymb}
 \usepackage{amsthm}

\journal{CSDA}\usepackage[utf8]{inputenc}
\usepackage{amsmath,amsfonts,amssymb,amsthm, multirow}
\usepackage{graphicx}
\usepackage{fullpage}
\usepackage{hyperref}
\usepackage{natbib}
\usepackage{color}
\usepackage{tabularx}
\usepackage{amsfonts,amssymb,amsmath,amscd,latexsym,dsfont}

\usepackage{graphicx}
\usepackage{subfigure}
\usepackage{natbib}
\newcommand{\balph}{\boldsymbol{\alpha}}
\newcommand{\bomega}{\boldsymbol{\omega}}

\newcommand{\bth}{\boldsymbol{\theta}}
\newcommand{\balpha}{\boldsymbol{\alpha}}

\newcommand{\bnu}{\boldsymbol{\nu}}

\newcommand{\bpi}{\boldsymbol{\pi}}
 \newcommand{\bm}{\boldsymbol{m}}

 \newcommand{\MM}{\mathcal{M}}
 
\newcommand{\bx}{\boldsymbol{x}} \newcommand{\tx}{\textbf{x}}
\newcommand{\tz}{\textbf{z}} \newcommand{\bz}{\boldsymbol{z}}
\newcommand{\ttt}{\textbf{t}} 
\newcommand{\MICL}{\text{MICL}}
\newcommand{\pdf}{p}
\newtheorem{ex}{Example}{\it}{\rm}

\newcommand{\tzstar}{\textbf{z}^{\star}}

\newcommand{\argmax}{\mathop{\mathrm{arg\,max}}}

\begin{document}

\begin{frontmatter}



\title{A tractable Multi-Partitions Clustering}

\author[label1]{Matthieu Marbac}
\author[label2,label3]{Vincent Vandewalle}

\address[label1]{CREST and Ensai}
\address[label2]{Univ. Lille, EA2694 Santé publique: épidémiologie et qualité des soins, F-59000 Lille, France}
\address[label3]{Inria}

\begin{abstract}
In the framework of model-based clustering, a model allowing several latent class variables is proposed. 
This model assumes that the distribution of the observed data can be factorized into several independent blocks of variables. 
Each block is assumed to follow a latent class model ({\it i.e.,} mixture with conditional independence assumption). 
The proposed model includes variable selection, as a special case, and is able to cope with the mixed-data setting. 
The simplicity of the model allows to estimate the repartition of the variables into blocks and the mixture parameters simultaneously, thus avoiding to run EM algorithms for each possible repartition of variables into blocks. 
For the proposed method, a model is defined by the number of blocks, the number of clusters inside each block and the repartition of variables into block. Model selection can be done with two information criteria, the BIC and the MICL, for which an efficient optimization is proposed. 
The performances of the model are investigated on simulated and real data. 
It is shown that the proposed method gives a rich interpretation of the dataset at hand ({\it i.e.,} analysis of the repartition of the variables into blocks and analysis of the clusters produced by each block of variables). 
\end{abstract}


\end{frontmatter}

\section{Introduction}

We consider the problem of multivariate data clustering. In this framework, an important issue is the choice of the variables used in the analysis. This choice can be performed  according to some focuses with respect to the desired clustering. Alternatively, without any prior knowledge on data, the statistician may perform the clustering based on all the available variables. Classical clustering methods assume that the considered variables explain a single partition among the observations. However, the available data could convey more that one partition of the data. For instance, one can imagine that different blocks of variables describing a customer (variables about work, variables about leisures, variables about family, \ldots) can give different clustering/partitioning of the dataset at hand. In absence of prior knowledge on how to group the variables into blocks, a challenging question for the statistician is to find these blocks of variables based on the data.

The problem of finding several partitions in the data, based on different groups of continuous variables, has been addressed by \cite{GALIMBERTI2007520} in a model-based clustering framework~\cite{McL00}. In this framework, the authors assume that the vector of variables can be partitioned in independent sub-vectors, each one following a particular mixture model. Then, they proposed a forward/backward search to perform model selection based on the maximization of the BIC. More recently,  \citet{Galimberti2017} have proposed an extension of their previous works which relaxes the independence assumption between sub-vectors. This extension considers three types of variables, the classifying variables, the redundant variables with respect to the classifying variables, and the variables which are not classifying at all. This can be seen as extension of the models proposed by \citet{raftery2006} and \citet{maugis2009variable}, in the framework of variable selection in clustering. Again, model selection is achieved with a forward/backward algorithm. 
Model selection is a difficult challenge because complex distributions are often used to model the data.
Therefore, they have to used forward/backward algorithms to maximize the BIC. 
However, these algorithms are suboptimal since they only converge to a local optimum of the BIC. 
Moreover, they are based on comparison of the BIC between two models. Thus, they perform many calls of EM algorithm. Hence, these approach only can deal with a limited number of variables (typically less than 100).

In order 
to deal with large numbers of variables, we propose an extension of the approaches proposed by \cite{marbac2017variable} and \cite{marbac2017variable2},  in the framework of variable selection in clustering. 
The main idea is to use a more constrained model to be able to easily perform model selection. 
We assume that the distribution of the observed data can be factorized into several independent blocks of variables, each one following its own mixture distribution. The considered mixture distribution in a block is a latent class model ({\it i.e.,} each variable of a block is supposed to be independent of the others given the cluster variable associated to this block). The simplicity of the model allows to estimate the repartition of the variables into blocks and the mixture parameters simultaneously~\citep{marbac2017variable,marbac2017variable2}. We present a procedure for performing model selection (choice of the number of blocks, the
number of clusters inside each block and the repartition of variables into block) with the BIC~\citep{Schwarz:78} or the MICL~\citep{marbac2017variable}. The BIC enjoys consistency properties and does not require to define prior distributions. However, in the clustering framework, it tends to over-estimate the number of clusters, and for small sample sizes the asymptotic approximation on which it relies can be questionable. Thus, in the framework of variable selection, \cite{marbac2017variable} have proposed the MICL criterion derived from the ICL criterion~\citep{biernacki2000assessing}. This criterion takes into account the classification purpose by computing the maximum integrated completed likelihood. Moreover, it is expected to well behave for small sample sizes, because it avoids the asymptotic approximations of the integrated completed likelihood by performing an exact integration over the parameter space thanks to conjugated priors. Depending on the context, either BIC or MICL can be preferred. In the context of multiple partitions clustering, it is possible to simultaneously perform parameter estimation (resp. partition estimation) and model selection with the BIC (resp. MICL) criterion like in~\cite{marbac2017variable,marbac2017variable2}, thus avoiding to run EM algorithms for each repartition of variables into blocks. Note that the proposed model allows to deal with mixed-data as in \cite{marbac2017variable2}, and it also includes the variable selection as a special case. Moreover, the proposed model can give an answer to problem of clustering mixed data in which continuous variables are often expected to dominate the clustering process. Allowing several partitions the categorical are now able, is necessary, to form their own clustering structure.  

The outline of the paper is the following. In Section~\ref{sec:model}, we present the multiple partitions mixture model. In Section~\ref{sec:ML}, we present the EM algorithm used for the estimation of the parameters by maximum likelihood when the blocks are known. In Section~\ref{sec:BIC}, we present how the model search can be performed using the BIC criterion. In Section~\ref{sec:MICL}, we present how the model search can be performed using the BIC criterion. In Section~\ref{sec:experiments}, we show the interest of the proposed model on simulated and real data.  

\section{Multiple partitions mixture model} \label{sec:model}
\subsection{The model}

The considered data $\tx=(\bx_1,\ldots,\bx_n)$ are composed of $n$ observations $\bx_i=(x_{i1},\ldots,x_{id})$ where $\bx_i$ is a vector of mixed variables, \emph{i.e.,} each variable can be continuous, binary, count or categorical. Moreover, we denote by $\tx_{j} = (x_{1j},\ldots,x_{nj})$ the observed data for variable $j$. The observations are assumed to be identically and independently drawn from a multiple partitions model (MPM) which is now described.

The MPM assumes that the variables are grouped into $B$ independent blocks, this repartition being encoded by $\bomega=(\omega_{j};j=1,\ldots,d)$, where $\omega_{j}=b$ indicates that variable $j$ belongs to block $b$. The set $\Omega_b=\{j: \omega_{j}=b\}$ denotes the indexes of variables of block $b$, and $\bx_{i\{b\}}=(x_{ij};j\in\Omega_b)$ is the vector of observed variables of block $b$. Let $\bz_{ib}$ be the class associated to group $b$ of observation $i$, $\bz_{ib} = (z_{ib1},\ldots,z_{ibG_b})$ with $z_{ibg} = 1$ if observation $i$ belongs to group $g$ for block $b$ and $z_{ibg} = 0$ otherwise. Let $\mathcal{Z}_{G}$ be the set of the partitions of $n$ elements in $G$ clusters, the partition related to block $b$ denoted by $\tz_{b} = (\bz_{ib}, \ldots, \bz_{ib})$ belongs to $\mathcal{Z}_{G_b}$. Thus the multiple partition $\tz=(\tz_1,\ldots,\tz_B)$ related to model $\bm=(B, G_1,\ldots, G_B, \bomega)$ belongs to $\boldsymbol{\mathcal{Z}}_{\bm}$ where  $\boldsymbol{\mathcal{Z}}_{\bm} = {\bf \mathcal{Z}}_{G_1}\times\ldots\times\mathcal{Z}_{G_B}$. Moreover, MPM assumes that $\bx_{i\{b\}}$ follows a $G_b$-component mixture distribution assuming the independence between variables of block $b$ given the latent class variable $\bz_{ib}$. Thus, the probability distribution function (pdf) of $\bx_i$ is
\begin{equation}
  \pdf (\bx_i | \bm, \bth ) = \prod_{b=1}^B   \pdf (\bx_{i\{b\}} | \bm, \bth )  
  \text{ with }
   \pdf (\bx_{i\{b\}} | \bm, \bth ) = \sum_{g = 1}^{G_b} \pi_{bg} \prod_{j \in \Omega_b} \pdf(x_{ij} | \balpha_{jg}),
 \end{equation}
where $\bth=(\bpi, \balpha)$ groups the model parameters, $\bpi=(\pi_{bg};b=1,\ldots,B;g=1,\ldots,G_b)$ groups the proportions with $\pi_{bg}>0$ and $\sum_{g=1}^{G_b} \pi_{bg}=1$, $\balpha=(\balpha_{jg};j=1,\ldots,d;g=1,\ldots,G_{\omega_j})$ groups the parameters of the univariate distributions. The univariate margin of a component  for a continuous (respectively binary, count and categorical), denoted by $\pdf(x_{ij} | \balpha_{jg})$, is a Gaussian (Bernoulli, Poisson and multinomial) distribution with parameters $\balpha_{jg}$ \citep{Mou05}.
 
\begin{ex}
To illustrate this distribution, we consider $d=4$ continuous variables generated by MPM with $B=2$ blocks of two variables. The first two variables belong to block 1 and the last two variables belong to block 2, hence $\omega_1=\omega_2=1$ and $\omega_3=\omega_4=2$. Moreover, each block follows a bi-component Gaussian mixture (\emph{i.e.,} $G_b=2$) with equal proportions (\emph{i.e.,} $\pi_{bg}=1/2$), mean $\mu_{1j}=4$, $\mu_{2j}=-4$ and variance $\sigma_{gj}^2=1$. Figure~\ref{fig:ex} gives the bivariate scatter-plots of the observations. Colors indicate the component memberships of block 1, and symbols indicate the component memberships of block 2.
\begin{figure}[ht!]
\begin{center}
\includegraphics[scale=.4]{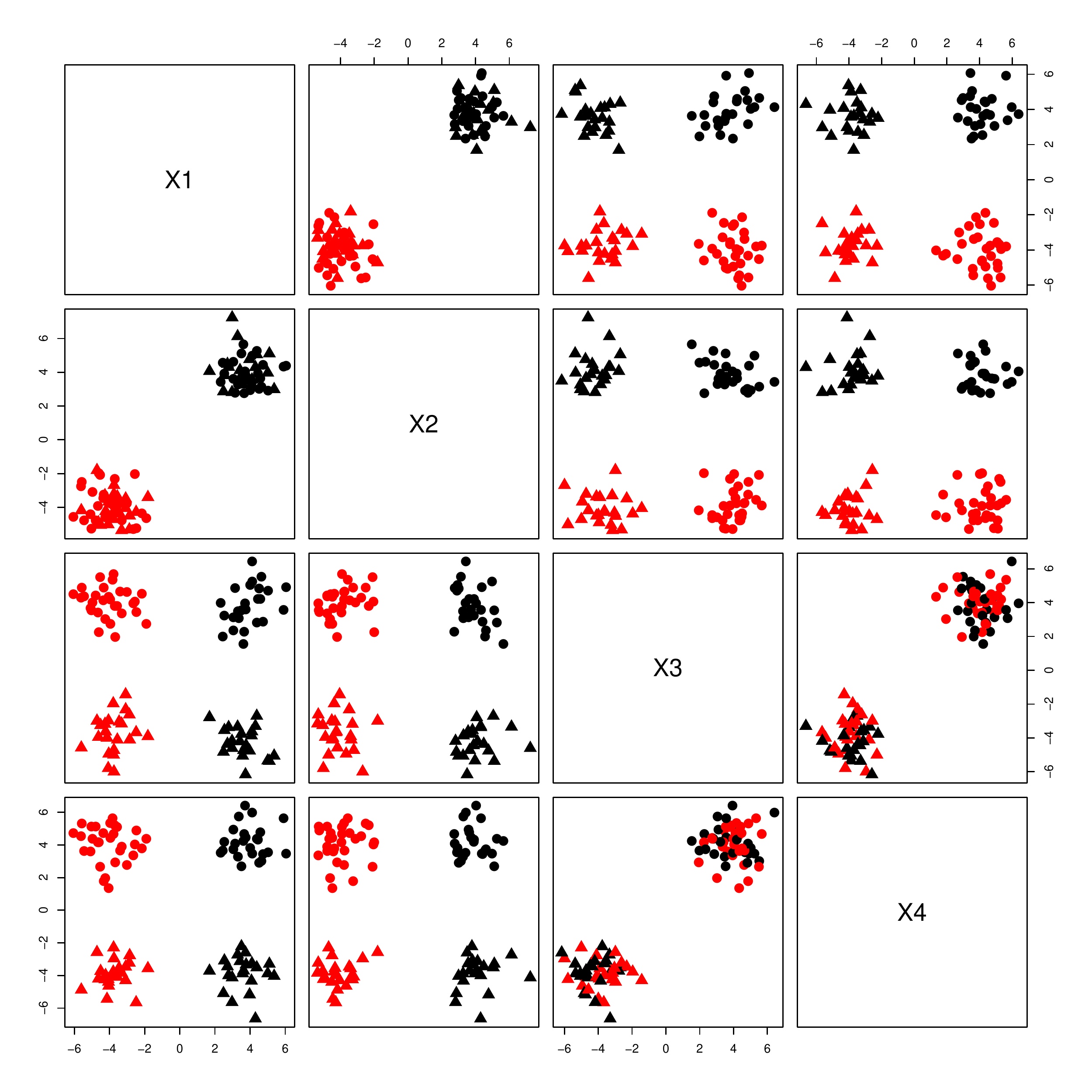}
\caption{Sample generated from MPM: colors indicate the partition of block~1 and symbol indicate the partition of block~2.\label{fig:ex}}
\end{center}
\end{figure}
\end{ex}

\subsection{Comments}
\paragraph{Link with approaches of model-based clustering}
Clustering approaches generally assume that there exists a single unobserved partition which is explained by the observed variables. However, this assumption can be wrong and MPM can be an answer to this problem. Indeed, this model  considers different partitions which are explained by subsets of variables.
Moreover, MPM generalizes approaches used for variable selection in model-based clustering. Indeed, if $B=2$ and $G_1=1$ then variables belonging to block 1 are not relevant for the clustering, while variables belonging to block 2 are relevant. Thus, MPM permits variable selection and multiple partitions explained by subsets of variables. 

\paragraph{Model identifiability}
In this paper, we consider that the vector of observations can be composed of variables with different natures. We consider the univariate distribution of the components for  continuous (respectively integer, categorical) variables are Gaussian (respectively Poisson, multinomial).
Model identifiability is directly obtained from the identifiability of Gaussian mixture with local independence \citep{Tei63, Tei67}.  Model identifiability requires that a block cannot be composed of two categorical variables. Indeed, if a block is composed by only categorical/binary variables, then the identifiability conditions of mixtures of multinomial distributions \citep{All09} must be validated. 

\paragraph{About the assumption of independence within components}
Finally, MPM assumes that variables are independent within components. This assumption permits to limit the number of parameters because model $\bm$ of MPM requires $\nu_{\bm}=\sum_{b=1}^B (G_b-1) + \sum_{j=1}^d card(\Theta_j) G_{\omega_j}$ parameters to be estimated, where $\Theta_j$ is the space of the parameters of the univariate margin of one component of variable $j$. Moreover, it permits efficient approaches for model selection (see Sections~\ref{sec:BIC} and \ref{sec:MICL}). However, the relaxation of this assumption is discussed in the conclusion.

\section{Maximum likelihood inference} \label{sec:ML}

For sample $\tx$ and model $\bm$, the observed-data log-likelihood is defined by
\begin{equation}
\ell(\bth|\bm,\tx)=\sum_{b=1}^B \sum_{i=1}^n \ln \pdf (\bx_{i\{b\}} | \bm, \bth )  .
\end{equation}
The model considers $B$ independent mixtures. 
The complete-data log-likelihood is
\begin{equation}
\ell(\bth|\bm,\tx,\tz)= \sum_{b=1}^B \ln p(\tz_b| \bpi_{b}) + \sum_{j=1}^d \ln p(\tx_{j}| \tz_{\omega_j}, \balpha_{j}), 
\end{equation}
where $\ln p(\tz_{b}| \bpi_{b})=\sum_{i=1}^n \sum_{g=1}^{G_b} z_{ibg} \ln \pi_{bg}$ and 
$ \ln p(\tx_{j}| \tz_{b}, \balpha_{j}) = \sum_{i=1}^n \sum_{g=1}^{G_b}  z_{ibg} \ln p(x_{ij}|\balpha_{jg})$. 
The maximum likelihood estimates (MLE) can be obtained by an EM algorithm \citep{Dem77,Mcl97}. Starting from the initial value $\bth^{[0]}$, its iteration $[r]$ is composed of two steps:\\
\textbf{E-step} Computation of the fuzzy partitions $t_{ibg}^{[r]}:=\mathbb{E}[Z_{ibg}|\bx_{i\{b\}},\bm,\bth^{[r-1]}]$, hence for $b=1,\ldots,B$, for $g=1,\ldots,G_b$, for $i=1,\ldots,n$
$$t_{ibg}^{[r]}=\dfrac{\pi_{bg}^{[r-1]} \prod_{j \in \Omega_b } \pdf(x_{ij}|\balpha_{jg}^{[r-1]})}{\sum_{k = 1}^{G_b} \pi_{bk}^{[r-1]} \prod_{j \in \Omega_b } \pdf(x_{ij}|\balpha_{jg}^{[r-1]})},$$
\textbf{M-step} Maximization of the expected value of the complete-data log-likelihood on $\bth$,
$$ \pi_{bg}^{[r]}=\dfrac{n_{bg}^{[r]}}{n} \text{ and } 
\balpha_{jg}^{[r]} = \argmax_{\balpha_{jg} \in \Theta_{j}} Q(\balpha_{jg}|\tx_{j},\ttt_{\omega_j g}^{[r]}),$$
where $Q(\balpha_{jg}|\tx_{j},\ttt_{b}) = \sum_{i=1}^n t_{ib g} \ln \pdf(x_{ij}|\balpha_{jg})$. 
Note that, independence between the $B$ blocks of variables permits to maximize the observed-data log-likelihood on each block independently. Thus, EM algorithms could be run on each block independently. This approach should be less sensitive to local optima. However, we choose to present the EM algorithm performing the maximization of the full observed-data likelihood, because this algorithm can be modified to perform the block estimation and the parameter inference simultaneously (see Section~\ref{sec:BIC}).

\section{Model selection with the BIC} \label{sec:BIC}
\subsection{Model selection}
Model have to be assessed from the data among a set of competing models $\MM$ defined by
\begin{equation}
\MM = \{ \bm: \omega_j \leq B_{\max} \text{ and } G_b \leq G_{\max}; j=1,\ldots,d; b=1,\ldots,B_{\max} \},
\end{equation}
where $B_{\max}$ is the maximum number of blocks and $G_{\max}$ is the maximum number of components within block. 
Model selection can be done by using the BIC \citep{Schwarz:78} defined by
\begin{equation}
\text{BIC}(\bm) = \max_{\bth_{\bm}} \ell_{pen}(\bth_{\bm}|\bm,\tx)
\end{equation}
where
\begin{equation}
 \ell_{pen}(\bth_{\bm}|\bm,\tx)
= \ell(\bth_{\bm}|\bm,\tx) - \frac{\nu_{\bm}}{2}\ln n,
\end{equation}
Model selection with the BIC consists in maximizing this criterion with respect to $\bm$. Obviously, this is equivalent to maximizing the penalized likelihood on the couple $(\bm,\bth_{\bm})$. Thus, model and parameter inference lead to search
\begin{equation}
(\bm^{\star}, \hat{\bth}_{\bm^{\star}}) = \argmax_{(\bm, \bth_{\bm})} \ell_{pen}(\bth_m|\bm,\tx).
\end{equation}

\subsection{Maximizing the penalized observed-data likelihood}
If $B$ and $(G_1,\ldots,G_B)$ are fixed, model selection with BIC and maximum likelihood inference imply to maximize the penalized likelihood on $(\bomega,\bth)$. In this section, we introduce a modified version of the EM algorithm \citep{green1990use} used for maximizing the penalized likelihood on  $(\bomega,\bth)$, for any $(B, G_1,\ldots,G_B)$. Thus, the combinatorial problem of model selection can be circumvented. Indeed, $(\bm^{\star}, \hat{\bth}_{\bm^{\star}})$ can be found by running this algorithm for each values of $B$ and $(G_1,\ldots,G_B)$ allowed by $\MM$. 
To implement this modified EM algorithm, we introduce the penalized complete-data likelihood
\begin{align}
\ell_{pen}(\bth_{\bm}|\bm,\tx,\tz)&= \ell(\bth_{\bm}|\bm,\tx,\tz) - \frac{\bnu_{\bm}}{2} \log n\\
&=  \sum_{b=1}^B \ln p(\tz_{b}| \bpi_{b}) - \frac{G_b-1}{2} \ln n  + \sum_{j=1}^d \ln p(\tx_{j}| \tz_{\omega_j}, \balpha_{j}) - \frac{\nu_{j} G_{\omega_j}}{2} \ln n,
\end{align}
where $\nu_j=\text{dim}(\Theta_j)$ (\emph{e.g.,} $\nu_j=2$ if the margin is a Gaussian distribution). 
Starting from the initial value $(\bomega^{[0]},\bth^{[0]})$, its iteration $[r]$ is composed of two steps:\\
\textbf{E-step} Computation of the fuzzy partitions $t_{ibg}^{[r]}:=\mathbb{E}[Z_{ibg}|\bx_i,\bm,\bth^{[r-1]}]$, hence for $b=1,\ldots,B$, for $g=1,\ldots,G_b$, for $i=1,\ldots,n$
$$t_{ibg}^{[r]}=\dfrac{\pi_{bg}^{[r-1]} \prod_{j \in \Omega_b^{[r-1]} } \pdf(x_{ij}|\balpha_{jg}^{[r-1]})}{\sum_{k = 1}^{G_b} \pi_{bk}^{[r-1]} \prod_{j \in \Omega_b^{[r-1]}} \pdf(x_{ij}|\balpha_{jg}^{[r-1]})},$$\textbf{M-step1} Updating the affectation of the variables to blocks
$$ \omega_{j}^{[r]}= \argmax_{\omega_{j}\in\{1,\ldots,B\}} \left(\sum_{g=1}^{G_{\omega_j}}\max_{\balpha_{jg} \in 
\Theta_j}  Q(\balpha_{jg}|\tx_{j},\ttt_{\omega_j g}^{[r]}) - \frac{\nu_j G_{\omega_{j}}}{2}\ln n \right),$$
\textbf{M-step2} Updating the model parameters
$$\pi_{bg}^{[r]}=\dfrac{n_{bg}^{[r]}}{n} \text{ and }
\balpha_{jg}^{[r]} = \argmax_{\balpha_{jg} \in \Theta_j} Q(\balpha_{jg}|\tx_{j},\ttt_{\omega_j^{[r]}g}^{[r]}).$$
Like for the classical EM algorithm, this modified EM algorithm converges into a local optimum of the objective function. Moreover, the objective function  increases at each iteration because $\ell_{pen}(\bth_m^{[r]}|\bm^{[r]},\tx)\geq\ell_{pen}(\bth_m^{[r-1]}|\bm^{[r-1]},\tx)$, with $\bm^{[r]}$ is the model defined by $(B,G_1,\dots,G_b)$ and $\bomega^{[r]}$. Thus, many random initializations should be done.

\section{Integrated complete-data likelihood}\label{sec:MICL}
\subsection{Model selection}
Criteria based on the integrated complete-data likelihood are popular for model-based clustering. Indeed, they take account into the clustering purpose (modeling the data distribution and providing well-separated components). Moreover, integrated complete-data likelihood has closed-form when components belong to exponential family and conjugate priors are used. The integrated complete-data likelihood is defined by
\begin{equation}
p(\tx,\tz|\bm) = \int p(\tx,\tz|\bm,\bth)p(\bth|\bm)d\bth.
\end{equation}
 We assume independence between the prior distributions, so $$p(\bth | \bm)  = \prod_{b=1}^B p(\bpi_b) \prod_{j \in \Omega_b} \prod_{g=1}^{G_b} p(\balpha_{jg}).$$ Thus, the integrated complete-data likelihood has the form defined by
\begin{eqnarray}
\ln p(\tx, \tz | \bm) &=& 
\sum_{b=1}^B \ln p(\tz_{b}|G_b) + 
 \sum_{j=1}^d \ln p(\tx_{j}|\tz_{\omega_j},G_{\omega_j},\omega_j) \\
 &=& \sum_{b=1}^B \left( \ln p(\tz_{b}|G_b) + 
 \ln p(\tx_{\{b\}}|\tz_{b},G_{b})\right),
\end{eqnarray}
where $p(\tz_{b}|G_b)=\int_{\mathcal{S}(G_b)} p(\tz_{b},\bpi_b|G_b)d\bpi_b$, $\mathcal{S}(G_b)$ denotes the simplex of dimension $G_b$ and  $ p(\tx_{j}|\tz_{\omega_j},G_{\omega_j},\omega_j)=\int_{\Theta_{j}^{G_{\omega_j}}}p(\tx_{j}|\tz_{\omega_j},\balpha_j, G_{\omega_j},\omega_j)p(\balpha_{j})d\balpha_j$. We use conjugate prior distributions. Thus, the integrals $p(\tz_{b}|G_b) $ and $ p(\tx_{j}|G_{\omega_j},\omega_j,\tz_{\omega_j})$  have closed forms (see Appendix~\ref{app:priors} for details).
 
The MICL (maximum integrated complete-data likelihood) criterion corresponds to the largest value of the integrated complete-data likelihood among all the possible partitions. Thus, the MICL is defined by
\begin{equation}
\MICL(\bm) = \ln p(\tx, \tzstar_{\bm} | \bm) \text{ with } \tzstar_{\bm}=\argmax_{\tz \in \boldsymbol{\mathcal{Z}}_{\bm}} \ln p(\tx, \tz | \bm).
\end{equation}
Model selection with MICL consists in finding the couple $(\bm^{\star},\tzstar_{\bm^{\star}})$ defined by
\begin{equation}
(\bm^{\star},\tzstar_{\bm^{\star}}) = \argmax_{(\bm,\tz)\in \mathcal{M} \times \boldsymbol{\mathcal{Z}}_{\bm}} p(\tx, \tzstar| \bm).
\end{equation}

\subsection{Maximizing the integrated complete-data likelihood}
For fixed number of block $B$ and numbers of components $G_1,\ldots,G_B$, maximizing MICL corresponds to maximizing the integrated complete-data likelihood on the affectation of the variables into block $\bomega$ and on the partition $\tz$. 
Starting at the initial value $\bomega^{[0]}$ where each $\omega_j$ is uniformly sampled among $\{1,\ldots,B\}$, the algorithm alternates between two steps defined at iteration $[r]$ by\\
\textbf{Partition step:} find $\tz_b^{[r]}$ such that for all $b = 1, \ldots, B$
$$p(\tx_{\{b\}}^{[r-1]}, \tz_b^{[r]}) \geq p(\tx_{\{b\}}^{[r-1]}, \tz_b^{[r-1]}),$$ 
where $\tx_{\{b\}}^{[r-1]} = (\tx_j ; \bomega^{[r-1]} = b)$.\\
\textbf{Model step:} find $\bomega^{[r]}$ such that for $j=1,\ldots,d$
$$\omega_j^{[r]} =\argmax_{b\in\{1,\ldots,B\}}  
 p(\tx_{j}|\tz_{b}^{[r]}).
$$
Optimization at the Partition step is not obvious, despite that it can be done on each block independently. So, the partition $\tz_b^{[r]}$ is defined as a partition which increases the value of the integrated complete-data likelihood for the current model for block $b$. It is obtained by an iterative method initialized with the partition $\tz_b^{[r-1]}$. Each iteration consists in sampling uniformly an individual which is affiliated to the class maximizing the integrated complete-data likelihood, while the other class memberships are unchanged. Optimization at the Model Step can be performed independently for each variable because of the intra-component independence assumption.
The optimization algorithm converges to a local optimum of the integrated complete-data likelihood. Thus, many different initializations should be done.

\section{Numerical experiments}\label{sec:experiments}
Numerical experiments are presented in this section. 
First, the performances of the method are investigated on simulated data. A robustness of the approach is illustrated by considering the within component dependencies. 
Second, the analysis of a mixed-data is conducted. 
Finally, we present the analysis of a challenging genomic data, where the number of variables is more than the number of observations. 

\subsection{Model performances on simulated data}
\paragraph{Simulation maps} In this section, we investigate the performances of the approach when the model is well-specified, then when the model is miss-specified (\emph{i.e.,} when variables are dependent within components). Thus, samples of size $n$ are generated from a model with three blocks (\emph{i.e.,} $B=3$) composed of one continuous variable and one integer variable each. The first two blocks follow bi-component mixture of Gaussian copulas (\emph{i.e.,} $G_1=G_2=1$) with the correlation coefficient $\rho$. For these two blocks, the univariate margin  of the continuous variable  for component $g$ follows a Gaussian distribution  with mean $\mu_{jg}=g \delta$ and variance $1$. For these two blocks, the univariate margin  of the integer for component $g$ follows a Poisson distributions with parameter  $g \delta$. The last block is composed of irrelevant variables (\emph{i.e.,} $G_3=1$) following a Gaussian copula with  correlation coefficient $\rho$. In this block, the univariate margin  of the continuous variable is a centered standard Gaussian distribution, and the univariate margin  of the integer variable  is a Poisson distribution with parameter $\delta$. For different values of $n$ (25, 50, 100, 200) and two values of $\rho$ (0 and 0.5), 25 replicates are sampled. Note that when $\rho=0$, the model is well-specified because variables are independent within components. Moreover, when $\rho=0.5$, the model is miss-specified because there are some dependencies within components. Finally, parameter $\delta$ is used for defining different overlaps between components. Thus the ``easy'' (resp. ``interm.'' and ``hard'') case corresponds to $\delta=4.5$ (resp. $\delta=3$ and $\delta=1.5$).

\paragraph{Results obtained when model is well-specified}
Table~\ref{tab:sim1} presents the results when the model is well-specified.
We note that, when the overlap between components is small, BIC and MICL behave identically. Even for small samples, they permit to detect the model (repartition of the variables and numbers of components). When the overlap between components increases (see interm. case), BIC obtains better results.  Indeed, MICL needs more observations than BIC to obtain the same results. Finally, when the overlap between components is high, MICL fails to detect the structure of the data. Indeed, because the entropy between components is too large,  MICL selects only one component. This results was expected, because criteria based on the complete-data likelihood can find the true model only when the component overlap is not too high.

\begin{table}[ht]
\centering
\begin{tabular}{cccccccc}
  \hline
 case & $n$ & \multicolumn{3}{c}{BIC} & \multicolumn{3}{c}{MICL}\\ 
& & $\hat{\bomega}$ & $\hat{G}$ & $\hat{\tz}$ & $\hat{\bomega}$ & $\hat{G}$ & $\hat{\tz}$\\ 
  \hline
easy & 25 & 0.80 & 0.64 & 0.90 & 0.64 & 0.68 & 0.83 \\ 
&  50 & 0.98 & 0.92 & 0.95 & 1.00 & 1.00 & 0.95 \\ 
&  100 & 0.93 & 1.00 & 0.98 & 0.98 & 1.00 & 0.98 \\ 
&  200 & 0.98 & 1.00 & 0.97 & 0.98 & 1.00 & 0.97 \\ 
interm. &  25 & 0.57 & 0.88 & 0.62 & 0.30 & 0.16 & 0.33 \\ 
&  50 & 0.71 & 0.72 & 0.66 & 0.53 & 0.32 & 0.45 \\ 
&  100 & 0.98 & 1.00 & 0.81 & 0.96 & 0.92 & 0.78 \\ 
&  200 & 0.98 & 1.00 & 0.82 & 0.98 & 1.00 & 0.82 \\ 
hard &  25 & 0.23 & 0.76 & 0.16 & -0.00 & 0.00 & 0.00 \\ 
&  50 & 0.18 & 0.96 & 0.14 & -0.00 & 0.00 & 0.00 \\ 
&  100 & 0.29 & 0.92 & 0.17 & 0.04 & 0.00 & 0.02 \\ 
 & 200 & 0.55 & 0.84 & 0.24 & 0.00 & 0.00 & 0.00 \\ 
   \hline
\end{tabular}
\caption{Results obtained by the BIC and the MICL when model is well-specified: ARI between the repartition of the variables into blocks and its estimate ($\hat{\bomega}$), frequency where the true vector of the number of components is found ($\hat{G}$) and ARI between the partitions and their estimates ($\hat{\tz}$).}\label{tab:sim1}
\end{table}

\paragraph{Results obtained when model is miss-specified}
Table~\ref{tab:sim2} presents the results when the model is miss-specified.
This simulation illustrates the robustness of MICL to the misspecification of the model because this criterion uses the component entropy. 
When the overlap between components is not too high, MICL detects the true repartition of the variables into blocks and the true numbers of components, while BIC fails to detect the true number of components. Indeed, BIC overestimates the number of components within blocks.

\begin{table}[ht]
\centering
\begin{tabular}{cccccccc}
  \hline
 case & $n$ & \multicolumn{3}{c}{BIC} & \multicolumn{3}{c}{MICL}\\ 
& & $\hat{\bomega}$ & $\hat{G}$ & $\hat{\tz}$ & $\hat{\bomega}$ & $\hat{G}$ & $\hat{\tz}$\\ 
  \hline
easy & 25  & 0.87 & 0.48 & 0.81 & 0.71 & 0.76 & 0.80 \\ 
 & 50 & 0.98 & 0.64 & 0.85 & 1.00 & 0.96 & 0.87 \\ 
 & 100 & 0.96 & 0.32 & 0.87 & 1.00 & 0.92 & 0.88 \\ 
 & 200 & 1.00 & 0.04 & 0.85 & 0.98 & 1.00 & 0.92 \\ 
interm. &  25 & 0.79 & 0.64 & 0.53 & 0.40 & 0.20 & 0.32 \\ 
  & 50 & 0.91 & 0.64 & 0.62 & 0.74 & 0.64 & 0.53 \\ 
 & 100 & 1.00 & 0.32 & 0.68 & 0.98 & 1.00 & 0.68 \\ 
 & 200 & 1.00 & 0.04 & 0.64 & 0.98 & 1.00 & 0.70 \\ 
hard & 25 & 0.57 & 0.76 & 0.19 & 0.25 & 0.08 & 0.10 \\ 
  & 50 & 0.93 & 0.60 & 0.24 & 0.23 & 0.00 & 0.07 \\ 
  & 100 & 1.00 & 0.44 & 0.25 & 0.33 & 0.20 & 0.12 \\ 
  & 200 & 1.00 & 0.04 & 0.28 & 0.51 & 0.28 & 0.18  \\
   \hline
\end{tabular}
\caption{Results obtained by the BIC and the MICL when model is miss-specified: ARI between the repartition of the variables into blocks and its estimate ($\hat{\bomega}$), frequency where the true vector of the number of components is found ($\hat{G}$) and ARI between the partitions and their estimates ($\hat{\tz}$).}\label{tab:sim2}
\end{table}

\begin{table}[ht]
\centering

\begin{tabular}{cccccccccc}
  \hline
 case & $n$ & \multicolumn{4}{c}{BIC results} & \multicolumn{4}{c}{MICL results} \\ 
& & $\hat{\bomega}$ & $\hat{G}$ & $\hat{\tz}_1$ & $\hat{\tz}_2$ & $\hat{\bomega}$ & $\hat{G}$ & $\hat{\tz}_1$ & $\hat{\tz}_2$\\ 
  \hline
easy & 25  & 0.26 & 0.12 & 0.79 & 0.96 & 0.23 & 0.76 & 0.84 & 0.99 \\ 
&50 & 0.52 & 0.04 & 0.90 & 0.97 & 0.40 & 0.44 & 0.87 & 0.96 \\ 
 &100 & 0.82 & 0.00 & 0.82 & 0.91 & 0.71 & 0.28 & 0.88 & 0.97 \\ 
  &200 & 0.93 & 0.00 & 0.72 & 0.78 & 0.87 & 0.04 & 0.83 & 0.92 \\ 
 interm. & 25 & 0.26 & 0.16 & 0.51 & 0.81 & 0.24 & 0.68 & 0.38 & 0.82 \\ 
 & 50 & 0.30 & 0.08 & 0.57 & 0.72 & 0.23 & 0.64 & 0.56 & 0.74 \\ 
 & 100 & 0.50 & 0.00 & 0.56 & 0.72 & 0.36 & 0.36 & 0.57 & 0.72 \\ 
 & 200 & 0.67 & 0.00 & 0.49 & 0.60 & 0.56 & 0.04 & 0.60 & 0.72 \\ 
hard & 25 & 0.16 & 0.48 & 0.04 & 0.18 & 0.07 & 0.20 & 0.00 & 0.07 \\ 
 & 50 & 0.30 & 0.24 & 0.05 & 0.15 & 0.15 & 0.32 & 0.01 & 0.07 \\ 
 & 100 & 0.47 & 0.08 & 0.08 & 0.19 & 0.28 & 0.60 & 0.01 & 0.09 \\ 
 & 200 & 0.65 & 0.00 & 0.11 & 0.18 & 0.50 & 0.72 & 0.02 & 0.08 \\ 
   \hline
\end{tabular}
\end{table}

\subsection{Contraceptive Method Choice data}

This dataset is a subset of the 1987 National Indonesia Contraceptive Prevalence Survey  \citep{lim2000comparison}. It describes 1473 Indian women with one continuous variable (AGE: age), one integer variable (Chi: number of children) and seven categorical variables (EL: education level, ELH: education level of the husband, Rel: religion, Oc: occupation, OcH: occupation of the husband, SLI: standard-of-living index and ME: media exposure).
Data are analyzed by considering at the maximum three blocks (\emph{i.e.,} $B_{\max}=3$) and six components (\emph{i.e.,} $G_{\max}=6$).	

The best three models according to the BIC are presented in Table~\ref{tab:appliBIC}. All of them consider two blocks. Moreover, the repartition of the variables into blocks is almost equal for these models. If we focus on the model selected by the BIC, we observe that the assumption of independence between blocks is relevant. Indeed, the Adjusted Rand Index computed on the partitions obtained by blocks 1 and 2 is equal to 0.01.
\begin{table}[ht!]
\begin{center}
\begin{tabular}{ccccccccc|cc|c}
\hline Age & Chi & EL & ELH & Rel & Oc & OcH & SLI & ME & $G_1$ & $G_2$ & BIC \\ 
\hline 
1 & 1 & 2 & 2 & 2 & 1 & 2 & 2 & 2 & 6 & 3 & -16078 \\ 
1 & 1 & 2 & 2 & 2 & 1 & 2 & 2 & 2 & 5 & 3 & -16081 \\ 
1 & 1 & 2 & 2 & 2 & 2 & 2 & 2 & 2 & 4 & 3 & -16088 \\ 
\hline 
\end{tabular} 
\caption{Best three models according to the BIC: block repartition, number of components per block and BIC values.\label{tab:appliBIC}}
\end{center}
\end{table}

If we analyse the results produced by the best model, we see on Figure~\ref{fig:contraception5}, that the two produced partitions seem rather uncorrelated, which is in accordance with the model assumptions.  
\begin{figure}[ht!]
\begin{center}
\includegraphics[width = 7cm]{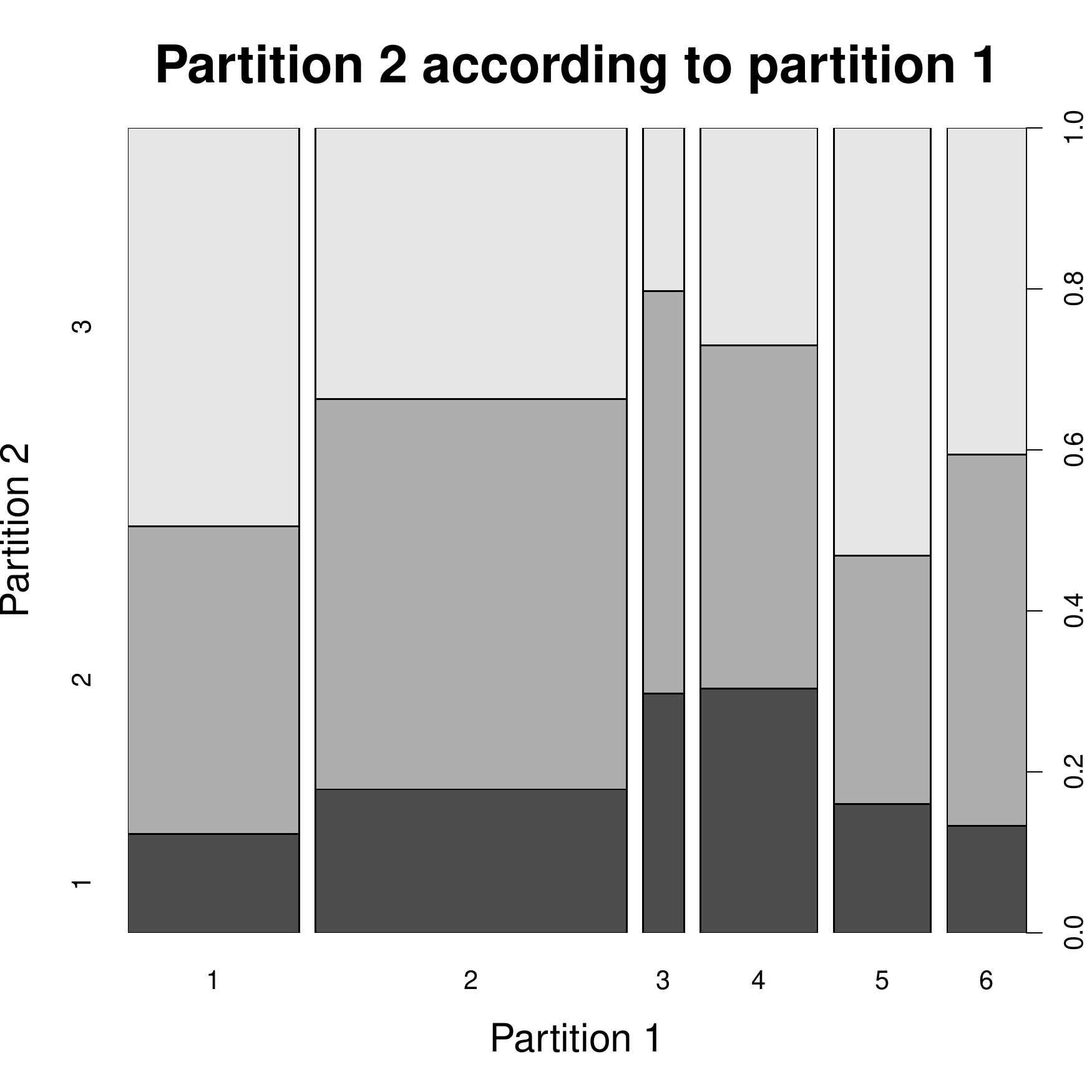}
\caption{\label{fig:contraception5} Distribution of the partition produced by the second block of variables given the partition produced by first block of variables.}
\end{center}
\end{figure}

\begin{figure}[ht!]
\begin{center}
\includegraphics[width = 15cm]{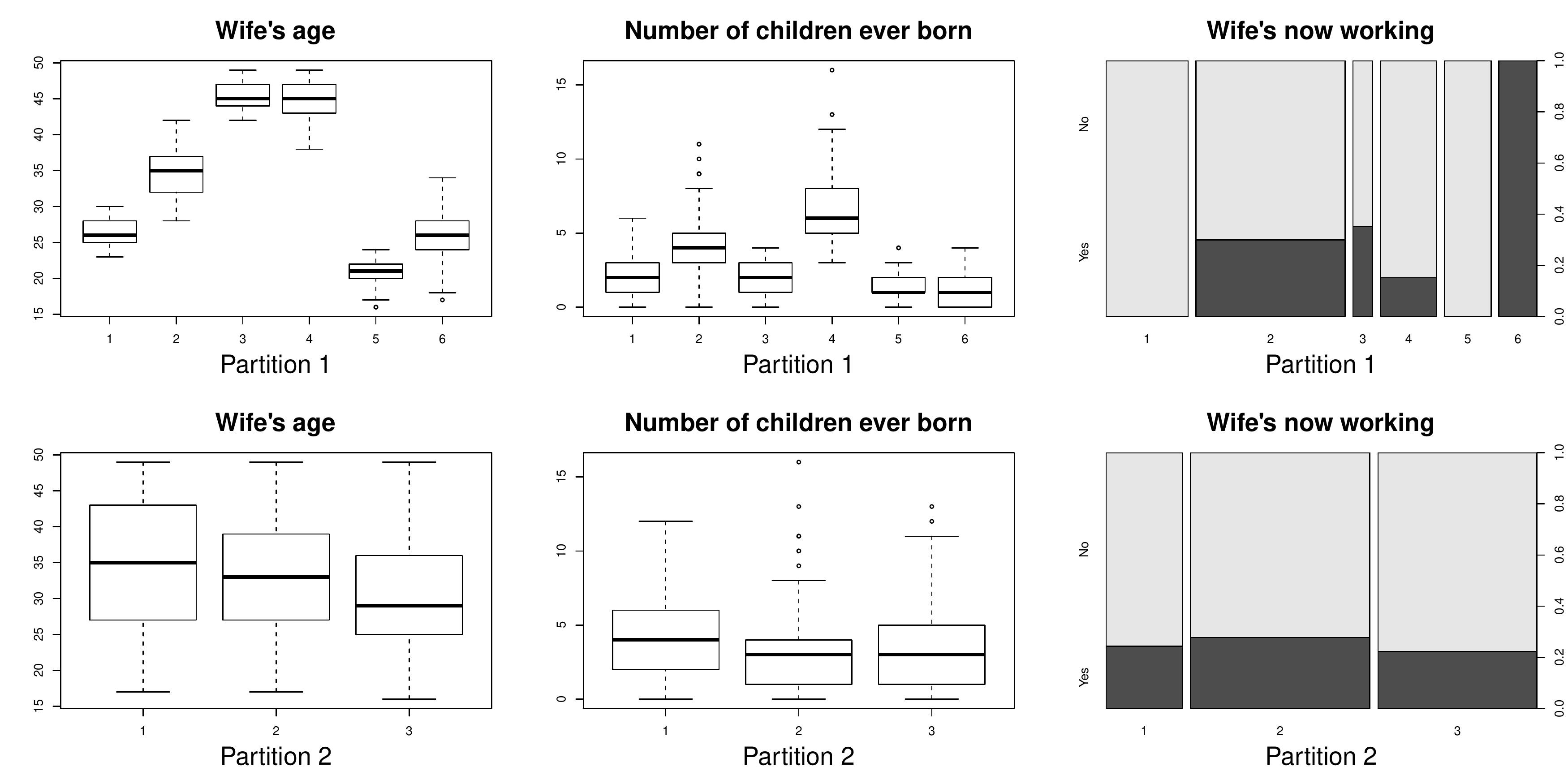}
\caption{\label{fig:contraception1} Distribution of the variables clustered in block 1 given the partition (partition 1 or partition 2).}
\end{center}
\end{figure}

On Figure~\ref{fig:contraception1}, we clearly see that the distributions of variables of block 1 (Wife's age, number of children and wife's now working) depend on the partition~1 while they seem rather independent of partition2.

\begin{figure}[ht!]
\begin{center}
\includegraphics[width = 15cm]{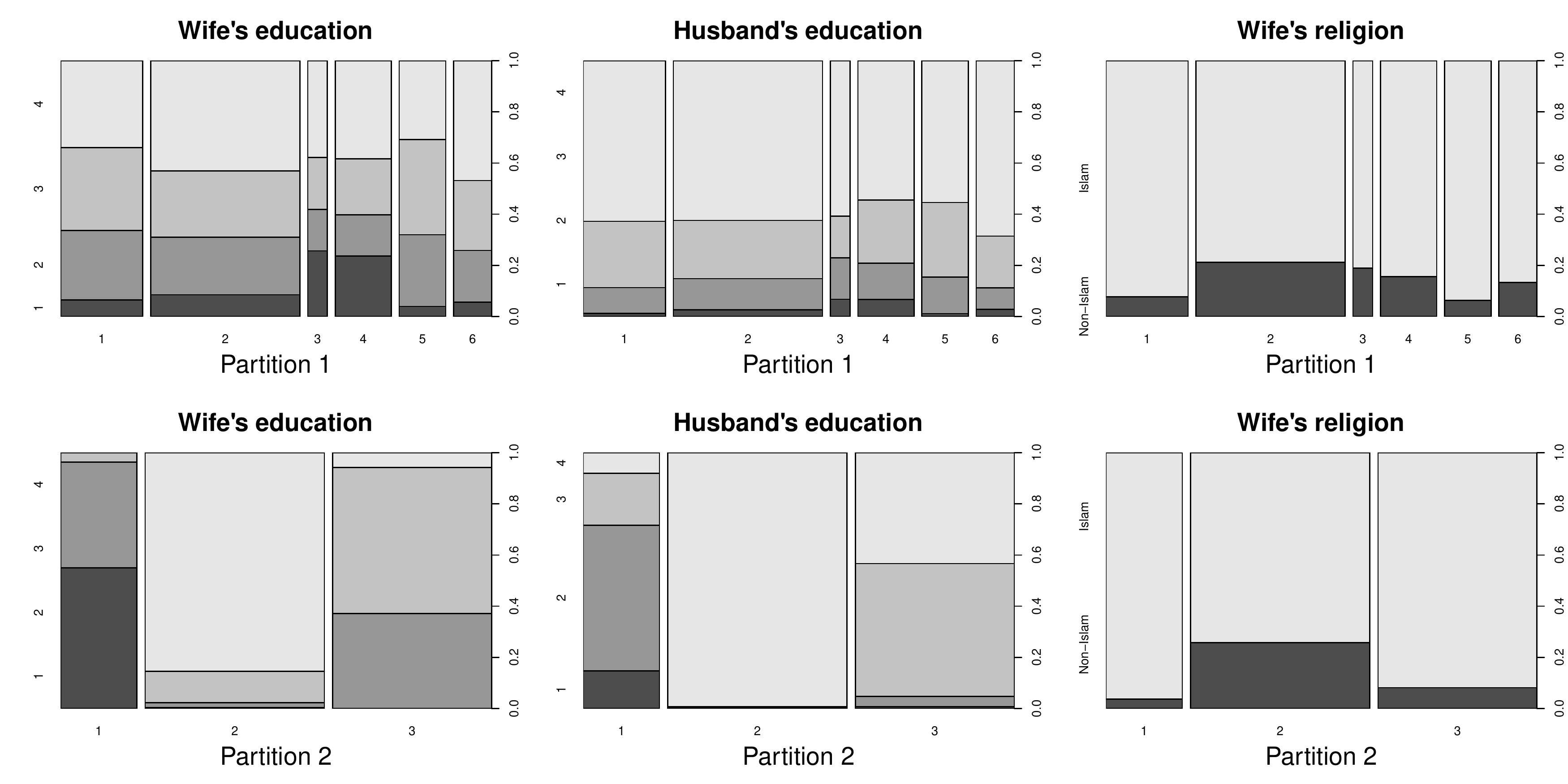}
\caption{\label{fig:contraception2} Distribution of the three first variables clustered in block 2 given the partition (partition 1 or partition 2).}
\end{center}
\end{figure}

\begin{figure}[ht!]
\begin{center}
\includegraphics[width = 15cm]{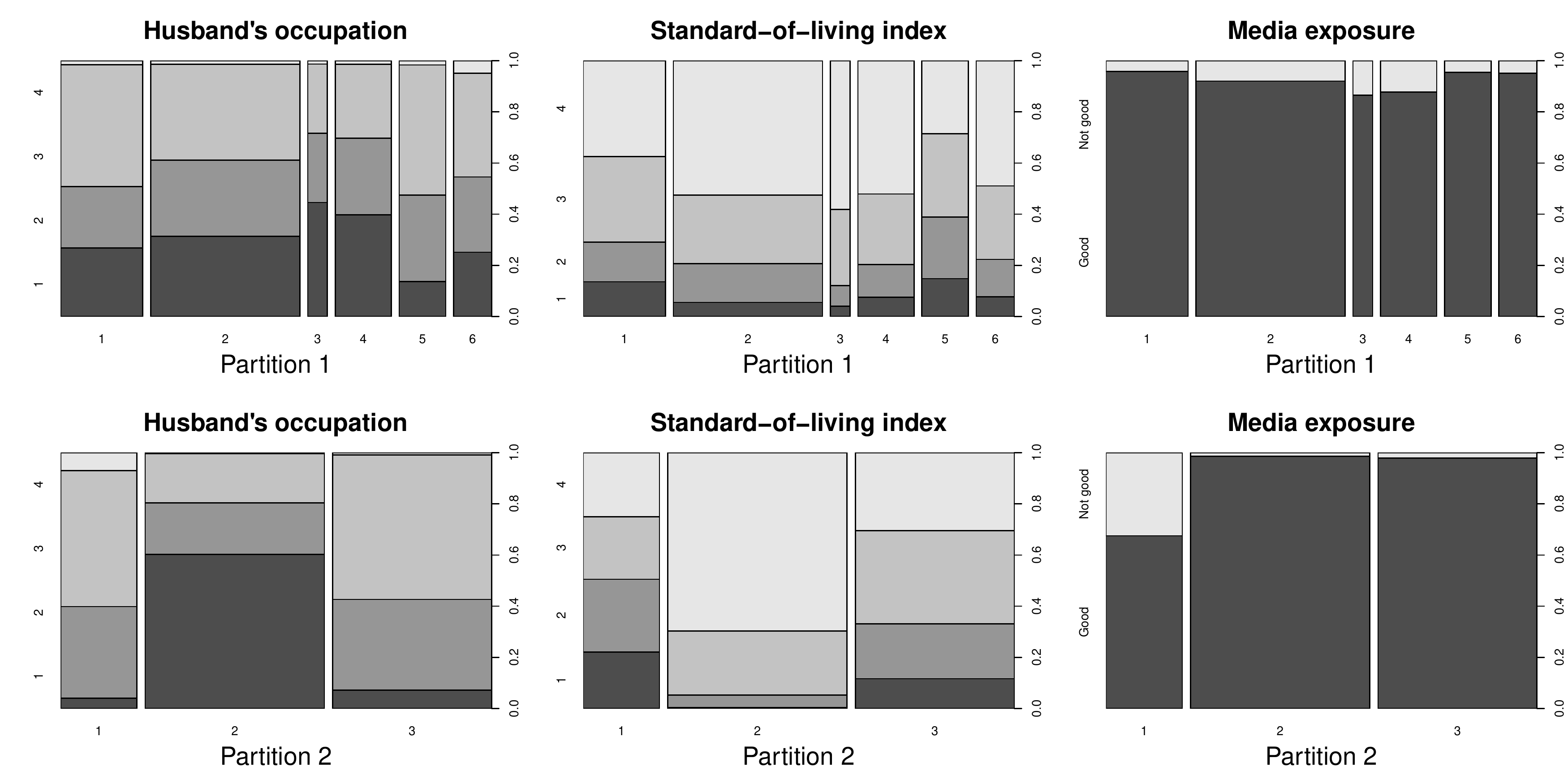}
\caption{\label{fig:contraception3} Distribution of the three last variables clustered in block 2 given the partition (partition 1 or partition 2).}
\end{center}
\end{figure}

On Figures~\ref{fig:contraception2} and~\ref{fig:contraception3}, we see that the distributions of variables of block 2  (Wife's education, husband's education, wife's religion, husband's occupation, standard-of-living index and media exposure)  depend on the partition 2 while they seem rather independent of partition 1.

\begin{figure}[ht!]
\begin{center}
\includegraphics[width = 10cm]{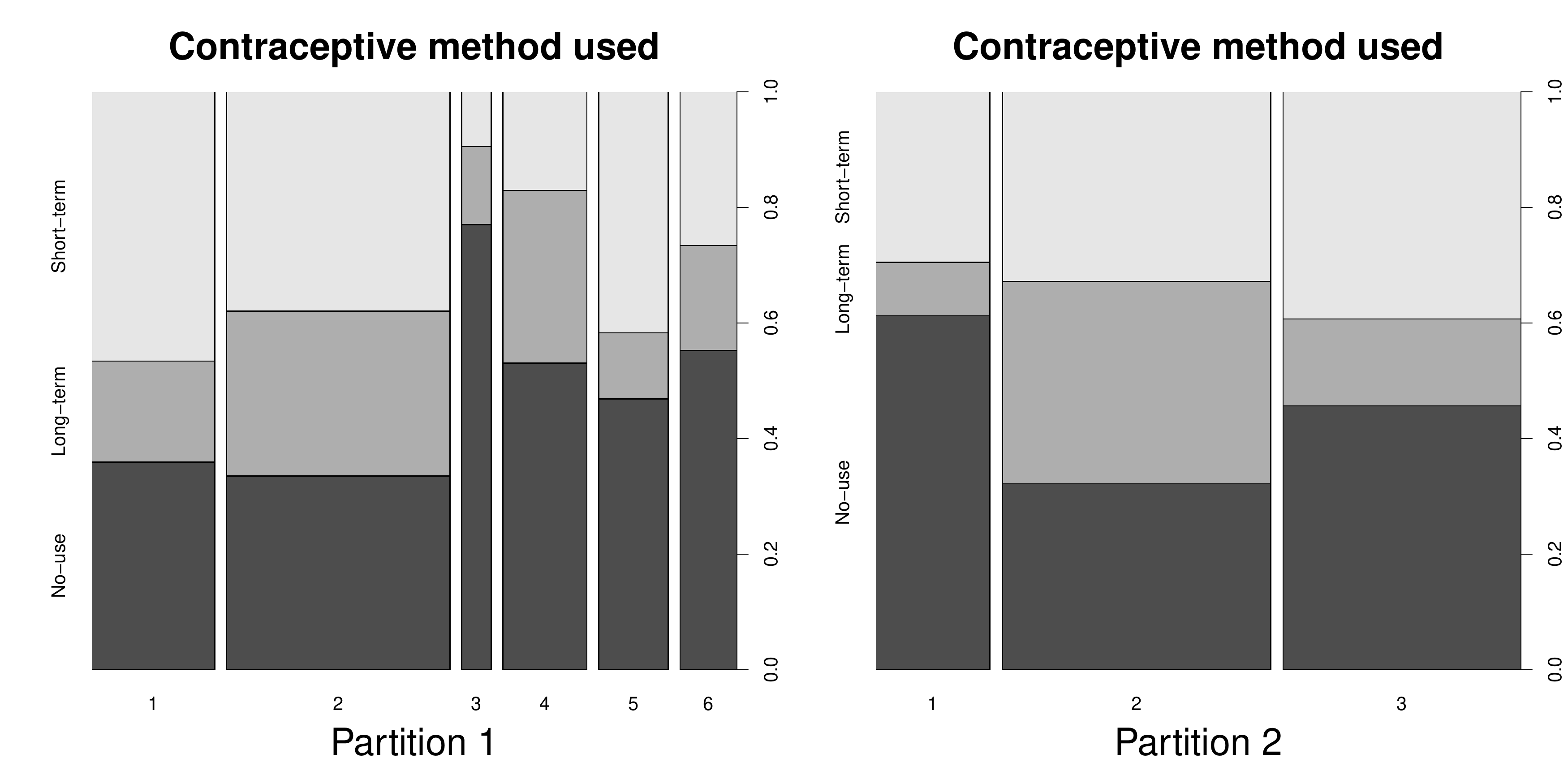}
\caption{\label{fig:contraception4} Distribution of the contraceptive method used given the partition (partition 1 or partition 2).}
\end{center}
\end{figure}

Finally, on Figure~\ref{fig:contraception4}, we study the distribution of the contraceptive methods used, which has not been used in the clustering and would be the target in the supervised classification setting. Here, the partition 1 seems the most correlated to the variable contraception, which could be expected because variables of block 1 are by definition linked with the contraceptive choice (Wife's age, number of children and wife's now working). Variables of block 2 rather produces a partition around the  the wife's education and the standard of living issue.   

We now consider the results obtained by the MICL. The best three models according to the MICL are presented in Table~\ref{tab:appliMICL}. The best two models select two blocks, while the third model selects only one block. For the best two models, the second block contains only one variable (Occupation). Because this block contains only a single component, this variable is detected as non relevant for the clustering. This show that the proposed approach permits to perform variable selection in clustering. 
\begin{table}[ht!]
\begin{center}
\begin{tabular}{ccccccccc|cc|c}
\hline Age & Chi & EL & ELH & Rel & Oc & OcH & SLI & ME & $G_1$ & $G_2$ & BIC \\ 
\hline 
1 & 1 & 1 & 1 & 1 & 2 & 1 & 1 & 1 & 4 & 1 & -16293 \\ 
1 & 1 & 1 & 1 & 1 & 2 & 1 & 1 & 1 & 5 & 1 & -16301 \\ 
1 & 1 & 1 & 1 & 1 & 1 & 1 & 1 & 1 & 4 & . & -16307 \\ 
\hline 
\end{tabular} 
\caption{Best three models according to the MICL: block repartition, number of components per block and MICL values.\label{tab:appliMICL}}
\end{center}
\end{table}

On Figure~\ref{fig:contraception5} we compare the partition produced by MICL with the other partitions previously studied. 

\begin{figure}[ht!]
\begin{center}
\includegraphics[width = 10cm]{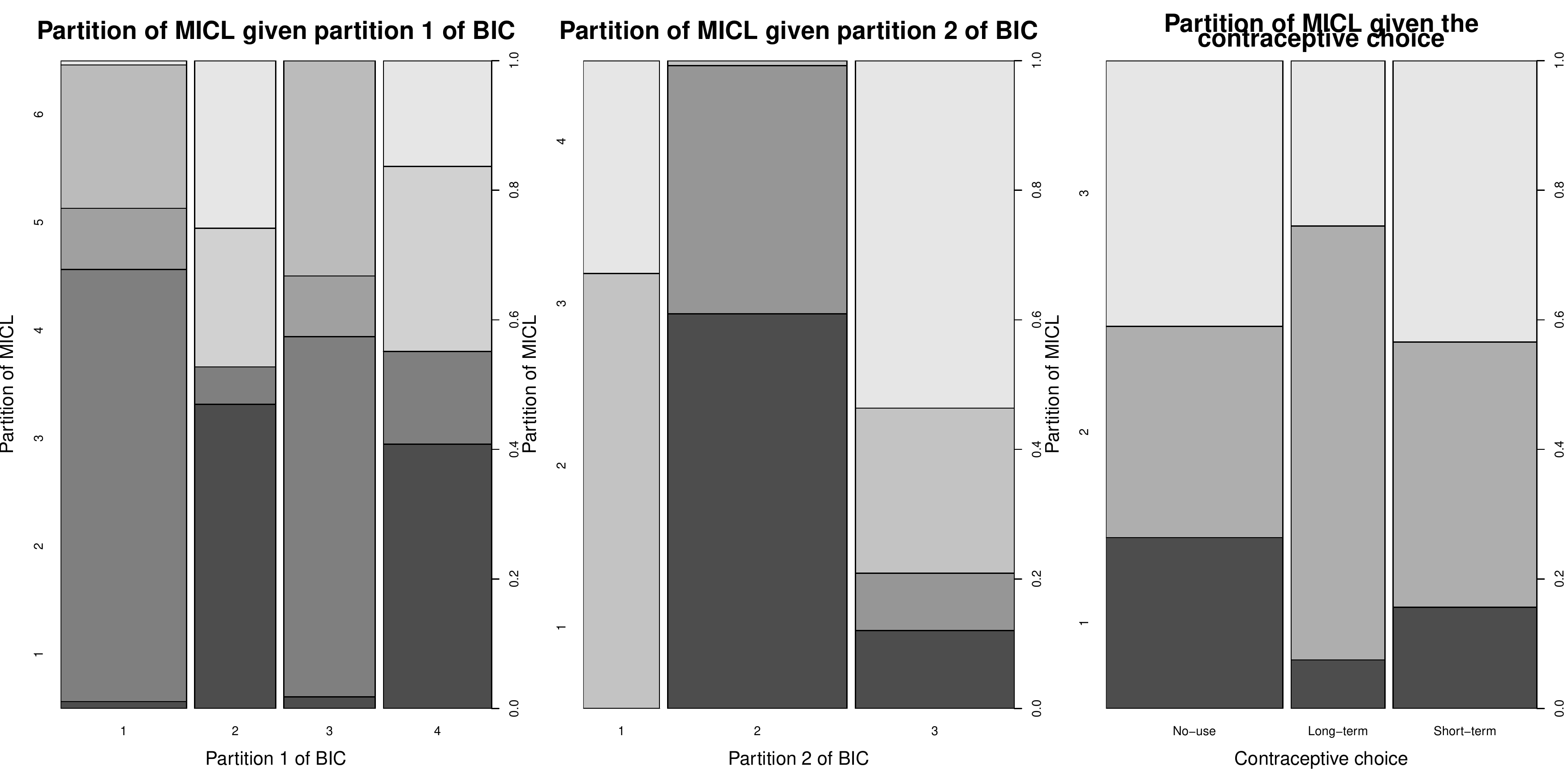}
\caption{\label{fig:contraception5} Distribution of the clustering produced by MICL given the two clustering produced by BIC and the contraceptive choice.}
\end{center}
\end{figure}


\subsection{Golub data}
We consider the dataset published in 1999 by \citet{golub1999molecular}. It showed how new cases of cancer could be classified by gene expression monitoring (via DNA micro-array) and thereby provided a general approach for identifying new cancer classes and assigning tumors to known classes. These data were used to classify patients with acute myeloid leukemia (AML) and acute lymphoblastic leukemia (ALL). This data describes $n=38$ patients with $d=3051$ continuous variables. Cluster analysis of such data are complex because the number of variables is more than the number of observations. In such cases, variable selection is especially important. We used the proposed approach to cluster this dataset. Note that the information about the type of cancer is hidden during the cluster analysis. This information is used to evaluated the performances of the clustering.

First, data are analyzed using a classical Gaussian mixture model (\emph{i.e.,} $B=1$ and $G_1\in\{1,\ldots,6\}$). Then, data are analyzed using a Gaussian mixture model performing variable selection (\emph{i.e.,} $B=2$, $G_1\in\{1,\ldots,6\}$ and $G_2=1$). Finally, data are analyzed with the bi-partition clustering approach performing variable selection  (\emph{i.e.,} $B=3$, $G_1\in\{1,\ldots,6\}$, $G_2\in\{1,\ldots,6\}$ and $G_3=1$).

Table~\ref{tab:Gol1} presents the results obtained by the three approaches when model selection is done with the BIC. The Gaussian mixture model performing variables selection permits to obtain a better value of the BIC than the classical Gaussian mixture. However, the resulting partition is less similar to the partition of reference. Thus, the $32\%$ relevant variables can explain an other structure between observations. The use of the bi-partition clustering approach performing variable selection is also relevant. First, note that this latter approach obtains the best value of the BIC. Moreover, the partition of block $1$ and the partition of reference are similar. Finally, block $2$ permits to detect an other structure among observations.

\begin{table}[ht!]
\begin{center}
\begin{tabular}{cccc}
\hline
B & G & ARI & $\%$ of variables\\
\hline 1 & $G_1=2$ &  0.70 & 100 \\  
\hline
2 & $G_1=4$  &  0.19 & 32 \\
& $G_2=1$ & 0.00 & 68 \\
\hline 3 & $G_1=3$ &  0.51 & 28\\
 & $G_2=4$ & 0.17 & 18 \\
 & $G_3=1$ & 0.00 & 54\\
 \hline
\end{tabular}
\end{center}
\caption{Results obtained with the BIC on Golub data: number of blocks ($B$), number of components per blocks ($G$), adjusted Rand index between the estimated partitions and the partition of reference (ARI), percentile of variables within blocks ($\%$ of variables).} \label{tab:Gol1}
\end{table}

For the Golub data clustering, well-separated clusters could be wanted. Moreover, the use of the BIC when $n<d$ can suffer from criticisms. Thus, analysis is also done with the MICL.
Table~\ref{tab:Gol2} presents the results obtained by the three approaches when model selection is done with the MICL.

\begin{table}[ht!]
\begin{center}
\begin{tabular}{cccc}
\hline
B & G & ARI & $\%$ of variables\\
\hline 1 & $G_1=1$ &  0.00 & 100 \\  
\hline
2 & $G_1=2$  &  0.79 & 18 \\
& $G_2=1$ & 0.00 & 82 \\
\hline 3 & $G_1=2$ &  0.70 & 16\\
 & $G_2=4$ & 0.51 & 6 \\
 & $G_3=1$ & 0.00 & 78\\
 \hline
\end{tabular}
\end{center}
\caption{Results obtained with the MICL on Golub data: number of blocks ($B$), number of components per blocks ($G$), adjusted Rand index between the estimated partitions and the partition of reference (ARI), percentile of variables within blocks ($\%$ of variables).} \label{tab:Gol2}
\end{table}

\begin{figure}[ht!]
\begin{center}
\includegraphics[scale=0.4]{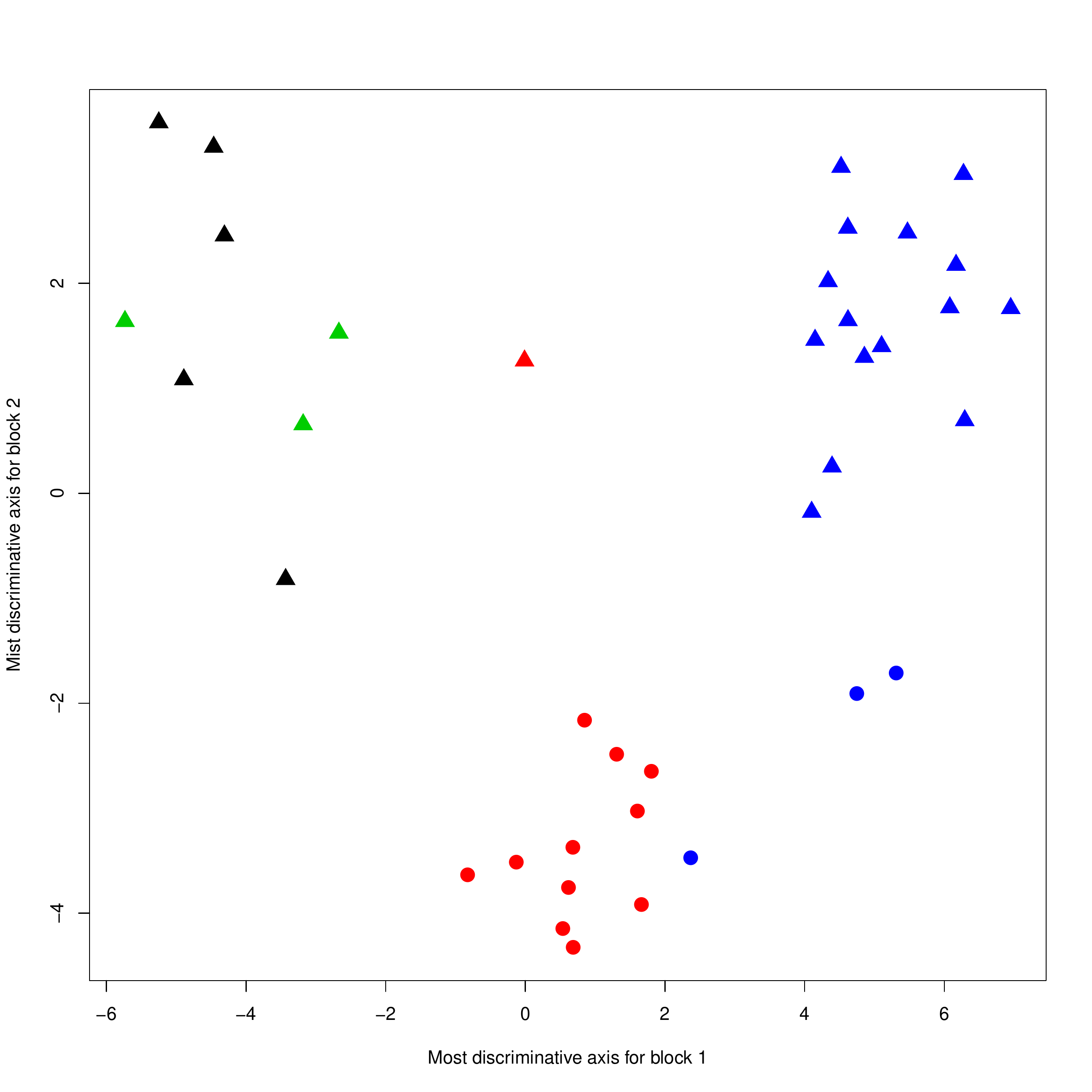}
\end{center}
\caption{Visualization of the multi-partitions clustering results: abscissa axis corresponds to the most discriminative axis for block 1 and ordinate axis corresponds to the most discriminative axis for block 2. Colors (resp. symbols) indicate the memberships of components of block~1 (resp. block 2).\label{fig:golub}}
\end{figure}

The best model according to MICL is defined by $B=3$ and $G=(2,4,1)$. The first block of variables is composed with $16\%$ of the observed variables and its partition is equal to the partition provided by the bi-component Gaussian mixture. Thus, it permits an easier interpretation of the partition, because this interpretation focuses only on a small subset of the variables. Moreover, the second block detects a specific structure defined by only $6\%$ of the observed variables. Finally, note that the Gaussian mixture performing variable selection provides the closest partition to the partition of reference.
To visualize the clustering results, a factorial discriminative analysis is performed on blocks 1 and 2. Figure~\ref{fig:golub} represents the observations on the map defined by the most discriminative axis of blocks 1 and 2. Colors (resp. symbols) indicate the memberships of components of block~1 (resp. block 2). Therefore, the abscissa axis permits to separate the colors while the ordinate  axis permits to separate the symbol. Note that the ordinate cannot discriminate the colors. This is in coherence with the assumption of independence between blocks.

This application illustrates that the multi-partition clustering with variable selection permits to detect different description on the variables. Moreover, interpretation is facilitated by variable selection. Indeed, the discriminative blocks (\emph{i.e.,} blocks $b$ with $G_b>1$) can be composed with a few number of variables. It makes no doubt that this relevant information can be hidden by a large number of non discriminative variables.

\section{Conclusion}

We have proposed a new method for performing clustering with multiple partitions. The proposed model is easily interpretable, and permits also to associate each produced partition with a subset of variables generating it. Thus, allowing to perform a clustering of variables of eventually different kinds as a by-product. Such kind of model allows in some sense to limit the subjectivity of the choice of variable in clustering, and allows to find several potentially interesting structures in the data without imposing that all the variables define the same clustering. The strength of the proposed approach is to use a simpler model, i.e. conditional independence assumption, than the state of the state of the art methods. Thus, the challenging problem of model selection can be circumvented, even for a large number of variables. Indeed, model selection can be done efficiently by maximizing classical information criteria (BIC or MICL).

The proposed method offers many possible extensions. On the first hand, since it performs  the clustering of the individuals and of the variables simultaneously, it can be in some sense interpreted as a co-clustering method. However to fit with the standard formulation of co-clustering with only one partition for the individuals, an additional modeling layer should be added to summarize the multi-partition by only a single partition. On the order hand, it would also be interesting in the quantitative setting to derive some k-means type approximation of the proposed method in order to deal with the very high dimensional setting as~\cite{witten2010framework} in the variable selection framework. 

\section*{Bibliography}

\bibliography{biblio}

\begin{thebibliography}{}

\bibitem[Allman et~al., 2009]{All09}
Allman, E., Matias, C., and Rhodes, J. (2009).
\newblock Identifiability of parameters in latent structure models with many
  observed variables.
\newblock {\em The Annals of Statistics}, 37(6A):3099--3132.

\bibitem[Biernacki et~al., 2000]{biernacki2000assessing}
Biernacki, C., Celeux, G., and Govaert, G. (2000).
\newblock Assessing a mixture model for clustering with the integrated
  completed likelihood.
\newblock {\em IEEE transactions on pattern analysis and machine intelligence},
  22(7):719--725.

\bibitem[Dempster et~al., 1977]{Dem77}
Dempster, A., Laird, N., and Rubin, D. (1977).
\newblock {Maximum likelihood from incomplete data via the EM algorithm}.
\newblock {\em Journal of the Royal Statistical Society. Series B
  (Methodological)}, 39(1):1--38.

\bibitem[Galimberti et~al., 2018]{Galimberti2017}
Galimberti, G., Manisi, A., and Soffritti, G. (2018).
\newblock Modelling the role of variables in model-based cluster analysis.
\newblock {\em Statistics and Computing}, 28(1):145--169.

\bibitem[Galimberti and Soffritti, 2007]{GALIMBERTI2007520}
Galimberti, G. and Soffritti, G. (2007).
\newblock Model-based methods to identify multiple cluster structures in a data
  set.
\newblock {\em Computational Statistics \& Data Analysis}, 52(1):520 -- 536.

\bibitem[Golub et~al., 1999]{golub1999molecular}
Golub, T.~R., Slonim, D.~K., Tamayo, P., Huard, C., Gaasenbeek, M., Mesirov,
  J.~P., Coller, H., Loh, M.~L., Downing, J.~R., Caligiuri, M.~A., et~al.
  (1999).
\newblock Molecular classification of cancer: class discovery and class
  prediction by gene expression monitoring.
\newblock {\em science}, 286(5439):531--537.

\bibitem[Green, 1990]{green1990use}
Green, P.~J. (1990).
\newblock On use of the em for penalized likelihood estimation.
\newblock {\em Journal of the Royal Statistical Society. Series B
  (Methodological)}, pages 443--452.

\bibitem[Lim et~al., 2000]{lim2000comparison}
Lim, T.-S., Loh, W.-Y., and Shih, Y.-S. (2000).
\newblock A comparison of prediction accuracy, complexity, and training time of
  thirty-three old and new classification algorithms.
\newblock {\em Machine learning}, 40(3):203--228.

\bibitem[Marbac and Sedki, 2017a]{marbac2017variable2}
Marbac, M. and Sedki, M. (2017a).
\newblock Variable selection for mixed data clustering: a model-based approach.
\newblock {\em arXiv preprint arXiv:1703.02293}.

\bibitem[Marbac and Sedki, 2017b]{marbac2017variable}
Marbac, M. and Sedki, M. (2017b).
\newblock Variable selection for model-based clustering using the integrated
  complete-data likelihood.
\newblock {\em Statistics and Computing}, 27(4):1049--1063.

\bibitem[Maugis et~al., 2009]{maugis2009variable}
Maugis, C., Celeux, G., and Martin-Magniette, M.-L. (2009).
\newblock Variable selection for clustering with gaussian mixture models.
\newblock {\em Biometrics}, 65(3):701--709.

\bibitem[McLachlan and Krishnan, 1997]{Mcl97}
McLachlan, G. and Krishnan, T. (1997).
\newblock {\em {The EM algorithm}}.
\newblock Wiley Series in Probability and Statistics: Applied Probability and
  Statistics, Wiley-Interscience, New York.

\bibitem[McLachlan and Peel, 2000]{McL00}
McLachlan, G. and Peel, D. (2000).
\newblock {\em Finite mixutre models}.
\newblock Wiley Series in Probability and Statistics: Apllied Probability and
  Statistics, Wiley-Interscience, New York.

\bibitem[Moustaki and Papageorgiou, 2005]{Mou05}
Moustaki, I. and Papageorgiou, I. (2005).
\newblock {Latent class models for mixed variables with applications in
  Archaeometry}.
\newblock {\em Computational statistics \& data analysis}, 48(3):659--675.

\bibitem[Raftery and Dean, 2006]{raftery2006}
Raftery, A.~E. and Dean, N. (2006).
\newblock Variable selection for model-based clustering.
\newblock {\em Journal of the American Statistical Association},
  101(473):168--178.

\bibitem[Schwarz, 1978]{Schwarz:78}
Schwarz, G. (1978).
\newblock Estimating the dimension of a model.
\newblock {\em The Annals of Statistics}, 6(2):461--464.

\bibitem[Teicher, 1963]{Tei63}
Teicher, H. (1963).
\newblock {Identifiability of Finite Mixtures}.
\newblock {\em The Annals of Mathematical Statistics}, pages 1265--1269.

\bibitem[Teicher, 1967]{Tei67}
Teicher, H. (1967).
\newblock Identifiability of mixtures of product measures.
\newblock {\em Annals of Mathematical Statistics}, 38:1300--1302.

\bibitem[Witten and Tibshirani, 2010]{witten2010framework}
Witten, D.~M. and Tibshirani, R. (2010).
\newblock A framework for feature selection in clustering.
\newblock {\em Journal of the American Statistical Association},
  105(490):713--726.

\end{thebibliography}
\bibliographystyle{apalike}

\appendix
\section{Closed-form of the integrated complete-data likelihood} \label{app:priors}
\subsection{Details about the prior distributions}
We use conjugate prior distributions, thus we assume that
\begin{itemize}
\item $\bpi_b|\bm$ follows a Dirichlet distribution $\mathcal{D}_{G_b}(u_1,\ldots,u_{G_b})$.
\item If variable $j$ is continuous, $\alpha_{jg}=(\mu_{jg},\sigma_{jg})$ where $\mu_{jg}$ is the mean of variable $j$ for component $g$ and $\sigma_{jg}$ is its standard deviation. We assume that $p(\balph_{jg})=p(\sigma_{jg}^2)p(\mu_{jg}|\sigma_{jg}^2)$ where $\sigma_{jg}^2$ follows an Inverse-Gamma distribution $\mathcal{IG}(a_j/2,b_j^2/2)$ and $\mu_{jg} | \bm, \sigma_{jg}^2$ follows a Gaussian distribution $\mathcal{N}(c_j, \sigma_{jg}^2/d_j)$.
\item If variable $j$ is integer,  $\balph_{jg}$ follows a Gamma distribution $\mathcal{G}a(a_j,b_j)$.
\item If variable $j$ is categorical, $\balph_{jg}$ follows a Dirichlet distribution $\mathcal{D}_{m_j}(a_{j},\ldots,a_{j})$ if variable $j$ is categorical with $m_j$ levels.
\end{itemize}
If there is no information \emph{a priori} on the parameters, we use the Jeffreys non-informative prior distributions  for the proportions (\emph{i.e.,} $u_{g}=1/2$) and for the hyper-parameters of a categorical variable (\emph{i.e.,} $a_{gj}=1/2$). 

\subsection{Details about the closed-form of the integrated complete-data likelihood}
To compute the  integrated complete-data log-likelihood, we give the values $p(\tz_{b}|G_b) $ and  $p(\tx_{j}|G_{\omega_j},\omega_j,\tz_{\omega_j})$ for the different types of data (continuous, integer and categorical).
\begin{itemize}
\item $p(\tz_{b}|G_b) = \frac{\Gamma\left(\frac{G_b}{2}\right)}{\Gamma\left(\frac{1}{2}\right)^{G_b}}\dfrac{\prod_{g=1}^{G_b} \Gamma\left(n_g + \frac{1}{2}\right)}{\Gamma\left(n+\frac{G_b}{2}\right)}$.
\item If variable $j$ is continuous, then
\begin{equation*}
p(\tx_{j}|\tz_{\omega_j}, G_{\omega_j},\omega_j)= 
\pi^{-n/2}
\left( \dfrac{b_j^{a_j/2}d_j^{1/2}}{\Gamma(a_j/2)}\right)^{G_{\omega_j}}
\prod_{g=1}^{G_{\omega_j}}
\dfrac{\Gamma(A_{gj}/2)}{B_{gj}^{A_{gj}}D_{gj}^{1/2}},
\end{equation*}
where $A_{gj}=n_{g\omega_j}+a_j$, $B_{gj}^2=b_j^2 + \sum_{i=1}^n z_{i\omega_jg} (x_{ij} - \bar{\text{x}}_{jk})^2 + \dfrac{(c_j - \bar{\text{x}}_{jg})^2}{d_j^{-1} + n_{g\omega_j}^{-1}}$, $D_{gj}=n_{g\omega_j} + d_j$, $\bar{\text{x}}_{jg}=\dfrac{1}{n_{g\omega_j}}\sum_{i=1}^n z_{i\omega_jg}x_{ij}$ and $n_{g\omega_j}=\sum_{i=1}^nz_{i\omega_jg}$.
\item If variable $j$ is integer, then
\begin{equation*}
p(\tx_{j}|\tz_{\omega_j}, G_{\omega_j},\omega_j)= 
\dfrac{1}{\prod_{i=1}^n \Gamma(x_{ij}+1)} \left( \dfrac{b_j^{a_j}}{\Gamma(a_j)}\right)^{G_{\omega_j}} \prod_{g=1}^{G_{\omega_j}} \Gamma(A_{gj})B_{gj}^{-A_{gj}}
, \end{equation*}
where $A_{gj}=\sum_{i=1}^n z_{i\omega_jg}x_{ij}+a_j$ and $B_j=b_j^2 + \sum_{i=1}^nz_{i\omega_jg}$.
\item If variable $j$ is categorical with $m_j$ levels, then
\begin{equation*}
p(\tx_{j}|\tz_{\omega_j}, G_{\omega_j},\omega_j)=
\left( \dfrac{\Gamma\big(m_j a\big)}{\Gamma(a)^{m_j}} \right)^{G_{\omega_j}}
\prod \limits_{g = 1}^{G_{\omega_j}}  
     \dfrac{\prod_{h = 1}^{m_j}\Gamma\big(\sum_{i=1}^n z_{i\omega_jg}\mathds{1}_{\{x_{ij}=h\}}  + a_j\big)}{\Gamma\big(\sum_{i=1}^n z_{i\omega_jg} + m_j a_j\big)} .
\end{equation*}
\end{itemize}

\end{document}